\documentclass[aps,prb,twocolumn,nofootinbib,superscriptaddress]{revtex4-2}
\usepackage{graphicx}
\usepackage{amsmath,amssymb}
\usepackage[utf8]{inputenc}
\usepackage[colorlinks = true,
linkcolor = blue,
urlcolor  = blue,
citecolor = blue,
anchorcolor = blue]{hyperref}
\usepackage{braket}

\usepackage[mathscr]{euscript}
\usepackage{mathdots}
\usepackage{color}
\usepackage{tikz}

\definecolor{lime}{HTML}{A6CE39}

\DeclareRobustCommand{\orcidicon}{%
\hspace{-1.0mm}
\begin{tikzpicture}
  \draw[lime, fill=lime] (0,0)
  circle [radius=0.15]
  node[white] {{\fontfamily{qag}\selectfont \tiny \,ID}};
  \draw[white, fill=white] (-0.0525,0.095)
  circle [radius=0.007];
\end{tikzpicture}
\hspace{-3.0mm}
}

\foreach \x in {A,...,Z}{%
  \expandafter\xdef\csname orcid\x\endcsname{%
    \noexpand\href{https://orcid.org/\csname orcidauthor\x\endcsname}%
    {\noexpand\orcidicon}}%
}

\begin{document}
\title{Second-order Skin Effect in a Brick-Wall Lattice}
\author{Dipendu Halder\orcidA{}}
\email[]{h.dipendu@iitg.ac.in} 
\author{Srijata Lahiri\orcidB{}}
\email[]{srijata.lahiri@iitg.ac.in}

\author{Saurabh Basu}
\email[]{saurabh@iitg.ac.in}
\affiliation{Department of Physics, Indian Institute of Technology Guwahati, Guwahati-781039, Assam, India}

\begin{abstract}
\noindent Non-Hermitian skin effect, which is a unique feature of non-Hermitian systems, exhibits the formation of an extensive number of boundary modes under open boundary conditions.
However, its manifestation in higher dimensions remains elusive.
In our work, we demonstrate a hybrid skin-topological effect arising from the interplay between first-order band topology and non-reciprocal hopping in an engineered two-dimensional brick-wall geometry.
The non-Hermitian brick-wall lattice under open boundary conditions in both directions exhibits several unconventional spectral features.
Notably, the eigenvalues associated with the corner skin modes do not exhibit non-trivial windings in the complex energy plane; instead, they exhibit dynamically stable exceptional point-like features that do not originate from eigenvector coalescence.
Of all the corner skin modes, only the two that originate from the topological corner states of the Hermitian brick-wall lattice remain localized at individual corners, while the rest accumulate at the pair of opposite corners.
This spatial distribution contrasts sharply with the second-order skin effect, where corner skin modes are more uniformly distributed.
Finally, for the non-Hermitian Brick-wall lattice, we design and implement the corresponding topolectrical circuit (circuit for a square lattice is included for comparison) to directly visualize the hybrid skin-topological modes.
\end{abstract}

\maketitle

\section{\label{Sec1}Introduction}

\noindent Two-dimensional lattices exhibiting Dirac-like crossings in their eigenspectra provide a versatile platform for studying topological quasiparticles and Berry phase effects on band topology \cite{RevModPhys.83.1057, PhysRevLett.100.096407} as well as emergent topological order \cite{PhysRevB.106.L121117}.
In this context, the honeycomb lattice provides a prominent platform for hosting Dirac Fermions where the effect of symmetry breaking on the topology of the hexagon has been extensively studied in connection with the anomalous Hall effect \cite{RevModPhys.82.1539}, $\mathbb{Z}_2$ \cite{PhysRevLett.95.146802} and Floquet topological insulators \cite{Lindner2011}, and several other related studies.
Evidently, Dirac-like band crossings present a useful setting for studying numerous emergent topological effects.
However, it is imperative to mention here, that such band crossings are not specific to the hexagonal lattice structure, but can be also obtained in other lattice geometries as well, including the Lieb and Kagome lattices \cite{PhysRevB.83.245125, PhysRevB.87.125428}.
Another model with an underlying square geometry, which is topologically equivalent to the honeycomb and consequently supports Dirac band crossings, is given by the Brickwall (BW) lattice \cite{PhysRevA.109.023322}.
Geometrically, the BW lattice can be obtained by stretching the honeycomb lattice appropriately so that the resulting structure resembles a square lattice with vertical bonds missing in alternate rungs.
The BW lattice is hence a two-dimensional bipartite structure that provides an alternative realization of the hexagonal topology, but with a much simpler lattice geometry.
This formulation has been employed to explore the topology of the hexagon and, hence, to study non-trivial insulating phases in cold atoms and optical lattices, owing to the ease of engineering a square lattice in artificial platforms \cite{PhysRevA.98.013603}.
Recent literature has also extensively explored the effect of hopping anisotropy, flux patterns and time-periodic perturbations that shift, merge or gap out Dirac-like band crossings in square and variants of the square lattices thereby generating non-trivial Berry curvature and novel emergent topological phases in a geometry much simpler than the hexagonal \cite{{doi:10.1126/sciadv.aat0346}, doi:10.1126/science.aah6442, w41n-yk21}.
Motivated by the above developments, we study a variant of the BW lattice that lacks an exact one-to-one correspondence with the stretched honeycomb structure. 
Our work primarily explores the effect of deviations from the square BW lattice to the conventional honeycomb structure, while maintaining the underlying rectangular lattice intact. In contrast, the hopping structure is altered to break the smooth connection to the hexagon.
The emergence of two Dirac cones is observed in the modified BW over a given region of parameter space, followed by their merger once a threshold condition is met.

Parallel to the above developments, non-Hermitian (NH) phenomena have experienced remarkable growth in recent years, finding applications across a wide range of condensed matter systems \cite{PhysRevLett.116.133903, PhysRevLett.120.146402, Ashida2020, RevModPhys.93.015005}. This rapidly developing field has unveiled a wealth of novel physical phenomena, such as the non-Hermitian skin effect (NHSE) \cite{PhysRevLett.121.086803, PhysRevB.99.201103, PhysRevLett.124.086801, PhysRevLett.124.056802, Zhang, Okuma}, where bulk eigenstates accumulate near the boundaries, and the emergence of exceptional points \cite{Heiss_2012, PhysRevLett.118.040401}, at which the Hamiltonian becomes defective due to the coalescence of eigenvalues and eigenvectors. In addition, the non-Bloch band theory \cite{PhysRevLett.121.086803, PhysRevLett.123.066404} has redefined the conventional Bloch theorem, providing new insights into the wave behavior of NH systems.
Experimental advances have further confirmed these phenomena across diverse physical platforms, including ultracold atoms \cite{Eichelkraut2013, El-Ganainy2018}, mechanical systems \cite{Wang}, photonic systems \cite{Weidemann2020}, and acoustic systems \cite{Fleury2015, 10.1063/5.0186638, 10.1063/5.0237506}. These developments have established NH systems as powerful platforms for exploring the interplay between topology and non-Hermiticity. Consequently, NH systems exhibit band structures with topological and localization properties that are fundamentally distinct from those of their Hermitian counterparts.

In purely Hermitian systems, higher-order topology arises from a non-trivial interplay between topological localization in two or more spatial directions, arranged so that the first-order topological polarizations cancel each other.
Consequently, in 2D systems, higher-order topology gives rise to $\mathcal{O}(1)$ zero modes localized at the corners, in stark contrast to the $\mathcal{O}(L)$ chiral or helical edge modes associated with first-order topology.
Likewise, in 3D systems, third-order topology produces $\mathcal{O}(1)$ zero modes confined to the corners, rather than the $\mathcal{O}(L^2)$ surface states characteristic of first-order TIs.
The advent of higher-order topology naturally motivates the study of higher-order skin effect~\cite{PhysRevB.102.205118, Zhang2022, PhysRevLett.123.016805, PhysRevB.99.081302}, particularly the second-order skin effect (SOSE) in NH analogues.
In 2D systems of size $L\times L$ with OBC along both directions, first-order NHSE is characterized by the accumulation of $\mathcal{O}(L^2)$ skin modes at generic boundaries.
By contrast, in the case of SOSE, only $\mathcal{O}(L)$ skin modes emerge, and these are localized at the corners.
In this regard, the hybrid-skin topological (HST) effect is noteworthy as it represents a distinct interplay between NHSE and conventional topological phases, resulting in eigenstate localization at lower-dimensional boundaries, such as corners in 2D lattices.
With the above motivation in mind, we embark on an NH variant of our modified BW model, in which non-reciprocal hopping along vertical bonds is the source of non-Hermiticity.

Although topological phenomena were originally discovered in condensed matter physics, they have since attracted strong interest from a broad range of fields, including photonics, phononics, mechanics, spintronics, and electrical circuits.
Among these platforms, electrical circuits have emerged as a compelling and accessible setting for investigating topological physics, introducing us to the concept of topolectrical circuits (TECs)~\cite{PhysRevLett.114.173902}.
Due to the high level of flexibility in connecting between two nodes, circuit networks have been used to create many novel states of matter, such as 2D topological states~\cite{Imhof2018, PhysRevLett.122.247702, PhysRevB.99.161114, Wu2022, Rafi-Ul-Islam_2020, PhysRevX.5.021031, PhysRevB.107.L201101}, 3D topological semimetals~\cite{Lee2018, 10.1093/nsr/nwaa192}, NH topological states~\cite{Helbig2020, PhysRevResearch.2.023265, Zou2021, PhysRevApplied.13.014047, PhysRevLett.124.046401, doi:10.34133/2021/5608038, PhysRevLett.126.215302, PhysRevB.98.201402}, higher-order topological states~\cite{Zhu2023, 10.1063/5.0157751, PhysRevLett.123.053902, PhysRevLett.124.036803, Yamada2022, Song2022}, and hyperbolic topological states~\cite{Lenggenhager2022, Chen2023, Zhang2023}.
The central idea underlying TECs is the exact correspondence between TB Hamiltonians in condensed matter systems and circuit Laplacians (admittance matrices).
In particular, circuit implementations offer substantial freedom in engineering hopping amplitudes and on-site potentials, including their strengths, directions, and dimensions of a TB model.
Moreover, TEC devices are well-suited for fabrication and miniaturization due to their compatibility with integrated circuit technology.
These advantages make TECs an effective platform for probing exotic topological phenomena, advancing the understanding of topological phases, and demonstrating potential practical applications.

The rest of the paper is organized as follows.
In Section~\ref{Sec2}, we first analyze the Hermitian BW lattice and discuss its band structure in momentum space.
Section~\ref{Sec3.1} analyzes the study of the SOSE in the NH square lattice.
In Section~\ref{Sec3.3}, we demonstrate how dynamically stable HST modes emerge when the NH square lattice is continuously transformed into the NH BW lattice.
For each case, we also propose corresponding TEC implementations for both the NH square lattice and the NH BW lattice (Sections~\ref{Sec3.2} and \ref{Sec3.4}), enabling direct visualization of the predicted HST modes.
Finally, Section~\ref{Sec4} summarizes the results.

\section{\label{Sec2}Model Hamiltonian}

\begin{figure}
\centering
\includegraphics[width=0.5\textwidth]{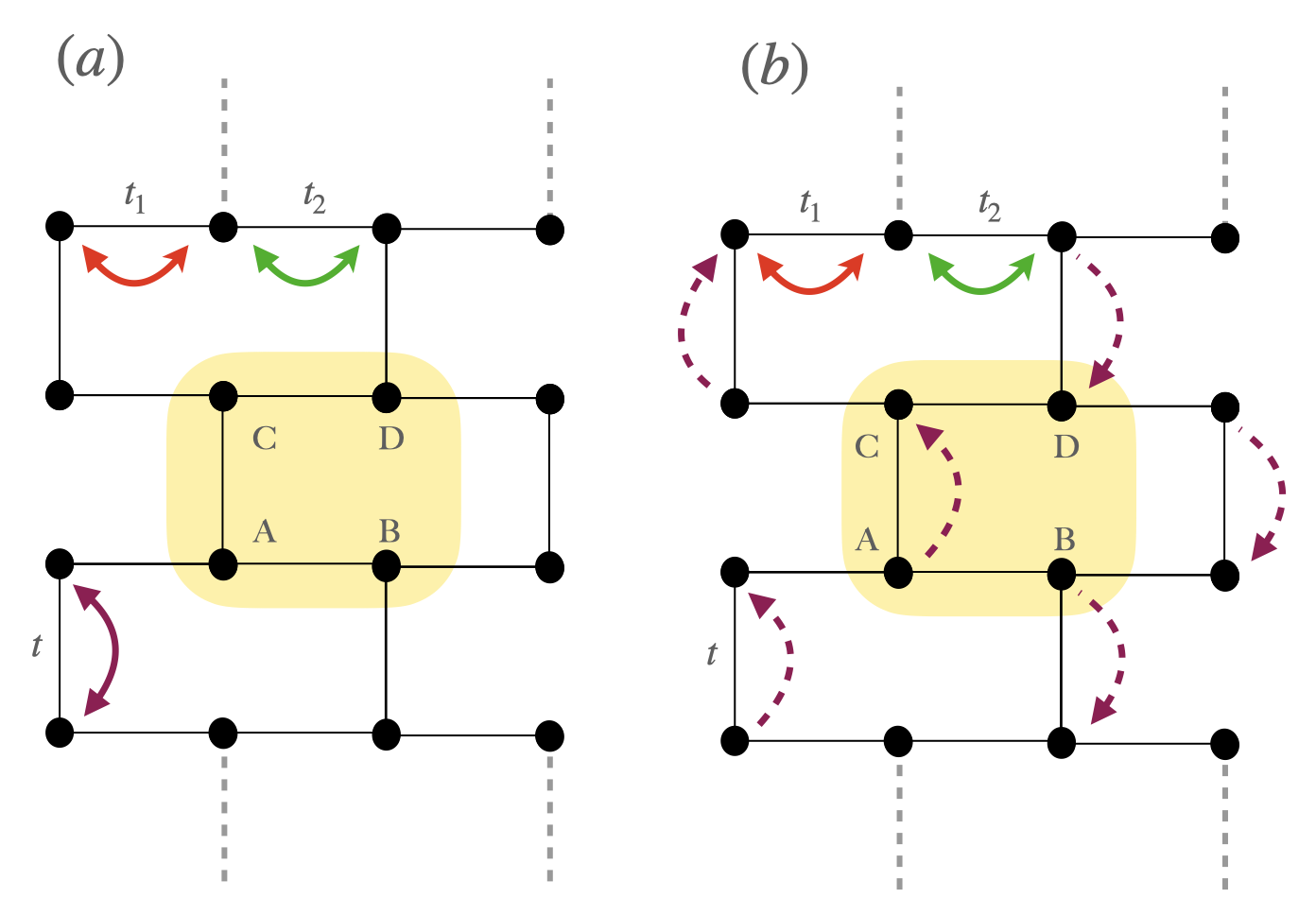}
\caption{(a) A schematic representation of the modified BW lattice is shown. The vertical hopping, which is only allowed between the intracell $A$ and $C$ sublattices and intercell $B$ and $D$ sublattices, is denoted by $t$. The horizontal direction resembles a stacked SSH model with a bipartite hopping structure characterized by hopping strengths $t_1$ and $t_2$. This model is Hermitian, with hopping allowed in both the forward and backward directions. (b) The NH version of the BW lattice, where the hoppings between $A\;(B)$ and $C\;(D)$ sublattices are unidirectional.}
\label{Fig1}
\end{figure}
\noindent The ordinary Hermitian BW lattice provides a convenient representation of a bipartite structure that is topologically similar to a honeycomb lattice, yet realized on a square (or rectangular) lattice.
The essential low-energy characteristics of the honeycomb lattice and, by extension, graphene are thus captured accurately in BW while remaining structurally straightforward.
In fact, the BW lattice can be visualized as a stretched honeycomb, with the hexagonal unit cell stretched to flatten the upper and lower vertices while preserving the hopping and sublattice structure. 
This modification retains the Dirac cones while changing the crystal symmetry from $C_6$ to $C_4$.
We study a slightly modified version of the original BW lattice, retaining the alternating vertical-hopping structure while discarding the direct correspondence with the honeycomb.
To begin with, our modified Hermitian BW consists of four sublattices, namely $A$, $B$, $C$, and $D$. 
The hopping in the vertical direction is denoted by $t$ and assumes an alternating structure similar to the original BW lattice (as shown in Fig.~\ref{Fig1}(a)).
To elaborate, while intra-cell hopping between the $A$ and $C$ sublattices is allowed and has a hopping amplitude $t$, inter-cell hopping between the sublattices is restricted, giving rise to the required BW structure.
In the horizontal direction, the hopping within the unit cell is denoted by $t_1$, whereas it is $t_2$ for the intra-unit cell hoppings.
The horizontal direction, therefore, resembles a stack of SSH chains with hopping amplitudes having a bipartite form, denoted by $t_1$ and $t_2$.
\begin{figure}
\centering
\includegraphics[width=0.5\textwidth]{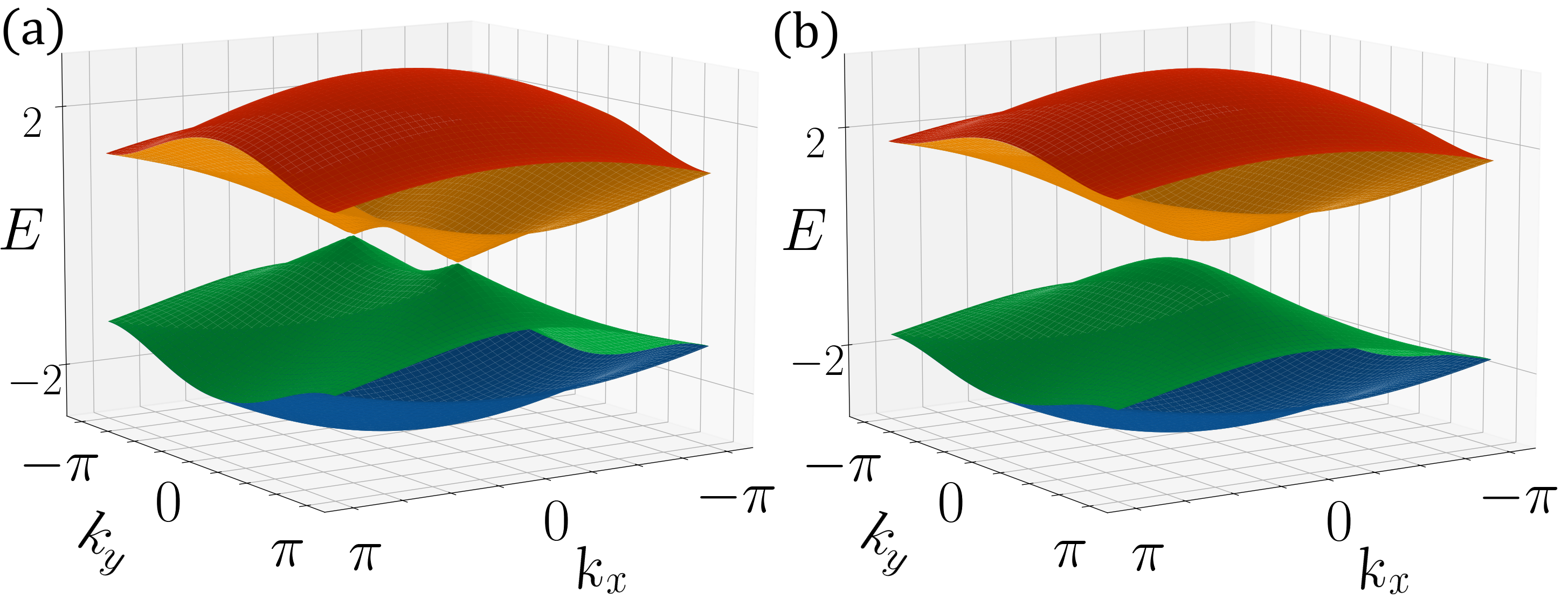}
\caption{The bulk bandstructure showing (a) the presence of four-fold Dirac cones for $t<t_1+t_2$ and (b) the gapped spectra for $t>t_1+t_2$.}
\label{Fig2}
\end{figure}
The modified BW lattice amalgamates the topology of the SSH model with the honeycomb to give rise to two four-fold degenerate Dirac cones in the regime $t<t_1+t_2$. 
At $t=t_1+t_2$, the Dirac cones merge at $\Gamma$ to finally gap out beyond this point.
The band structure is, however, completely invariant with respect to the relative magnitude of $t_1$ and $t_2$.
Having discussed the Hermitian BW model, we now move to the NH BW lattice, where we introduce non-reciprocal hopping in the vertical direction, which drastically changes the nature of the energy eigenspectra.

\section{\label{Sec3}Results and discussion}
\noindent As mentioned earlier, the modified Hermitian BW lattice exhibits the presence of two Dirac cones in the bulk Hamiltonian when the strength of the vertical hopping $t$ is less than the strength of the horizontal hopping $t_1$ and $t_2$ combined, as shown in Fig.~\ref{Fig2}(a).
\begin{figure}
\centering
\includegraphics[width=0.5\textwidth]{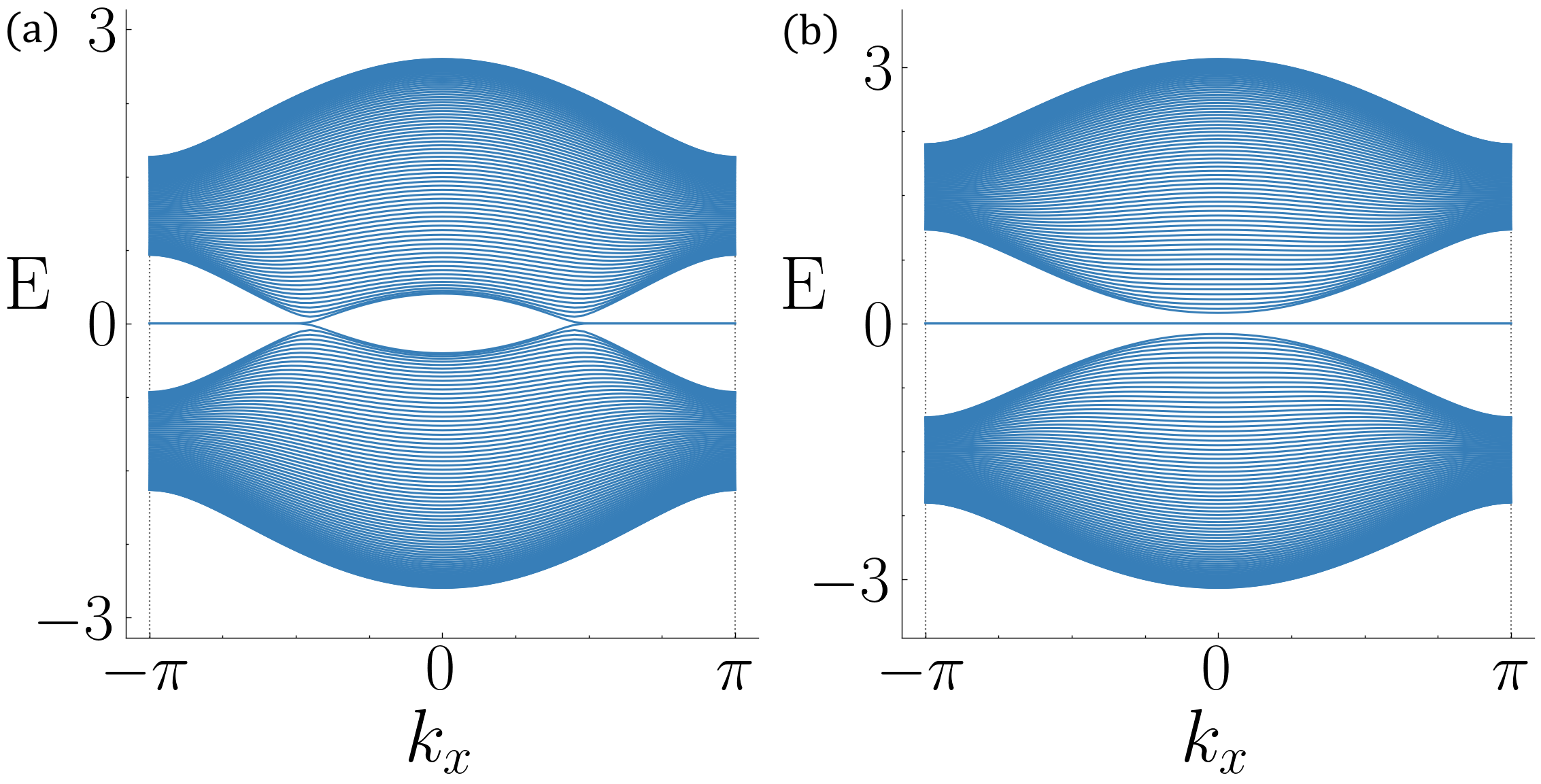}
\caption{The ribbon spectra for (a) $t<t_1+t_2$ and (b $t>t_1+t_2$. For $t>t_1+t_2$, the spectrum is completely gapped. However, the existence of an in-gap flat band is observed.}
\label{Fig3}
\end{figure}
The Dirac cones move towards each other and merge at the $\Gamma$ point when $t=t_1+t_2$, finally gapping out when $t>t_1+t_2$ (Fig.~\ref{Fig2}(b)). A nanoribbon geometry shows the Dirac points being connected by a flat band (Fig.~\ref{Fig3}(a)). For $t>t_1+t_2$, the valence and the conduction bands in the nanoribbon are completely disconnected. 
However, in-gap states are found to traverse the energy gap in this parameter region (Fig.~\ref{Fig3}(b)).
For the NH counterpart of the modified BW lattice, we introduce non-reciprocal hopping in the system such that two consecutive vertical rungs allow an upward hopping with strength $t$. In contrast, the next two vertical rungs exhibit downward hopping of the same magnitude.
We will return to the detailed consequences of this construction shortly.
Before doing so, however, it is instructive first to consider a reference system that exhibits a SOSE, and for that, we discuss NHSE in an NH square lattice.

\subsection{\label{Sec3.1}SOSE in an NH Square Lattice}

\noindent We begin with the simple Hermitian square lattice already discussed, in which the chains along the $x$-direction emulate the traditional SSH model, featuring staggered hopping amplitudes $t_1$ and $t_2$. At the same time, those along the $y$-direction are uniform with amplitude $t$.
In this particular variant of the model, non-reciprocity is introduced by rendering the $y$-direction chains non-reciprocal, allowing hopping only in one direction (see Fig.~\ref{Fig1}(b)).
The directions of the hopping are reversed along the alternate chains; that is, the hopping amplitudes along the nearest neighbor vertical chains are oriented in mutually opposite directions.
The bulk Hamiltonian for this model is given via,
\begin{align}
&H_s(\Vec{k})= \sum_{k_x,k_y} \Big[
\left(t_1+t_2e^{-ik_x}\right)c^{\dag}_{\Vec{k},A}c_{\Vec{k},B}\nonumber \\
&+ \left(t_1+t_2e^{ik_x}\right)c^{\dag}_{\Vec{k},C}c_{\Vec{k},D}+ te^{-ik_y}c^{\dag}_{\Vec{k},A}c_{\Vec{k},D} + te^{ik_y}c^{\dag}_{\Vec{k},C}c_{\Vec{k},B} \Big].
\label{eq:C7BSS}
\end{align}
$H_s(\vec{k})$ has both the mirror symmetries, $\mathcal{M}_x$ and $\mathcal{M}_y$, which are defined via,
\begin{equation}
    \mathcal{M}_x\equiv\tau_0\otimes\sigma_x;\quad\mathrm{and}\quad
    \mathcal{M}_y\equiv\tau_x\otimes\sigma_0,
\end{equation}
where $\tau$ and $\sigma$ are the Pauli matrices that describe additional degrees of freedom.
Consequently, $H_s(\vec{k})$  has the inversion symmetry, which is defined via the product of $\mathcal{M}_x$ and $\mathcal{M}_y$.
A closely related model to $H_s(\vec{k})$, that exhibits SOSE, was discussed by Kawabata {\it et al.} in Ref.~\cite{PhysRevB.102.205118}.
It was further argued that such an NH square lattice can be systematically extended to a Hermitian Benalcazar–Bernevig–Hughes (BBH) model, which serves as a paradigmatic example of a second-order topological insulator, originally proposed by Benalcazar, Bernevig, and Hughes~\cite{PhysRevB.96.245115, doi:10.1126/science.aah6442}.
Specifically, if $H(k)$ exhibits SOSE, then the corresponding Hermitian BBH Hamiltonian in $k$ space can be constructed as,
\begin{equation}
\tilde{H}_{\mathrm{BBH}}(k)=
\begin{pmatrix}
0 & H_s(k) \\
H_s^{\dagger}(k) & 0
\end{pmatrix}.
\label{eq:C7BBH}
\end{equation}
In this model, no edge modes appear when open boundary conditions are imposed along only one spatial direction.
However, when the system is open along both directions, zero-energy modes emerge at the corners once the intra-cell hopping becomes weaker than the inter-cell hopping.
This model respects both the inversion and the chiral symmetries, which together play a crucial role in stabilizing higher-order topology.
In particular, when a 2D Hermitian system possesses chiral and inversion symmetries, which are denoted by mutually anti-commuting operators, intrinsic second-order topology emerges~\cite{PhysRevB.97.205136}.
As a consequence, the Hamiltonian $H(k)$ supports an intrinsic SOSE, as proposed in Ref.~\cite{PhysRevB.102.241202}, where topological zero-energy corner modes are protected by the bulk topology.
By extending the notion of second-order topology to point gaps, the authors of Ref.~\cite{PhysRevB.102.241202} further demonstrated that NH systems can host skin modes localized at the corners.
This scenario is referred to as an intrinsic SOSE, as the corner localization is directly rooted in the bulk topological properties.
In contrast, an extrinsic SOSE arises when the bulk is topologically trivial, while the edges are topologically non-trivial.
In such cases, the edge bands carry an imprint of 1D first-order topology protected by the chiral symmetry, and the edge–corner correspondence yields corner modes through closing of the edge gap at zero energy.
The non-trivial topology of the 1D edges can then support skin modes localized at the corners.
However, unlike $H(k)$ in Eq.~\eqref{eq:C7BBH}, the NH square-lattice Hamiltonian $H_s(\vec{k})$  cannot be extended to a Hermitian Hamiltonian (with inversion symmetry) that exhibits intrinsic second-order topology.
As a result, $H_s(\vec{k})$  does not exhibit intrinsic SOSE.
Instead, it realizes an extrinsic SOSE, according to Ref.~\cite{PhysRevB.102.241202}, where the corner skin modes originate from the interplay between non-hermiticity and first-order topology of the SSH chains along the $x$-direction (see Fig.~\ref{Fig1}(a)), also referred to as the HST effect.
\begin{figure}
\centering
\includegraphics[width=0.5\textwidth]{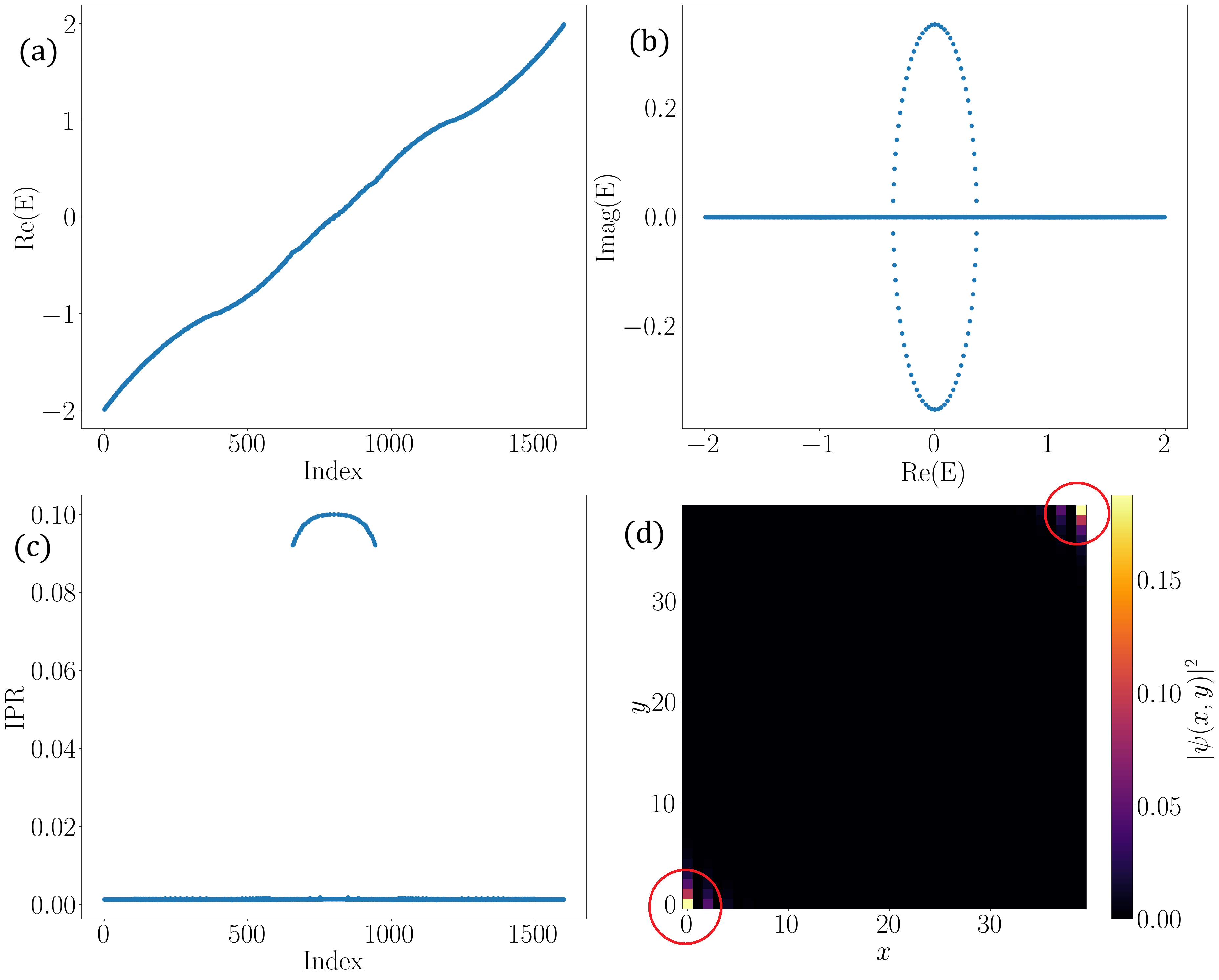}
\caption{(a) The real part of the energy spectra of a finite NH square lattice (OBC in both directions) for $t<t_1+t_2$. (b) The spectral distribution in the complex plane. The corner skin modes are arranged in the form of an ellipse. (c) IPR of the corner skin modes is non-zero. (d) The probability distribution of a random second-order skin mode, shown via yellow dots. For clarity, these dots are marked inside red circles.}
\label{Fig4}
\end{figure}

We now turn to the spectral and localization properties of the NH square lattice with open boundaries imposed along both directions, focusing on the regime $t<t_1+t_2$.
The real parts of the eigenenergies, shown in Fig.~\ref{Fig4}(a), indicate the absence of an energy gap.
The corresponding complex-energy spectra, plotted in Fig.~\ref{Fig4}(b), reveal that the spectral trajectories associated with the HST modes exhibit a non-trivial winding in the complex plane.
Such winding is forbidden in the case of the first-order NHSE, highlighting the distinct properties of these modes.
Fig.~\ref{Fig4}(c) displays the IPR, which is finite for the corner skin modes, demonstrating their localized character.
These modes are not part of the bulk spectrum and are instead confined to the system boundaries, with localization occurring specifically at the corners, while the remaining bulk modes remain delocalized.
Out of the total number of eigenstates ($L_x\times L_y$ of them), the number of corner skin modes is $2L_y$, whereas the number of delocalized bulk modes is $L_y\times(L_x-2)$.
This counting follows from the fact that each SSH chain along the $x$-direction contributes two zero-energy topological modes, and there are $L_y$ such chains in total.
For the numerical results presented here, we set $L_x=L_y=40$, yielding $80$ HST modes.
A representative probability distribution for one such corner skin mode is shown in Fig.~\ref{Fig4}(d).
Importantly, when the SSH chains along the $x$-direction are tuned into the topologically trivial regime, that is, $t_1>t_2$, these corner skin modes disappear.
This confirms that their existence originates from the interplay between first-order topology and non-hermiticity.

\subsection{\label{Sec3.2}TEC of an NH Square Lattice}

\begin{figure*}
\centering
\includegraphics[width=\textwidth]{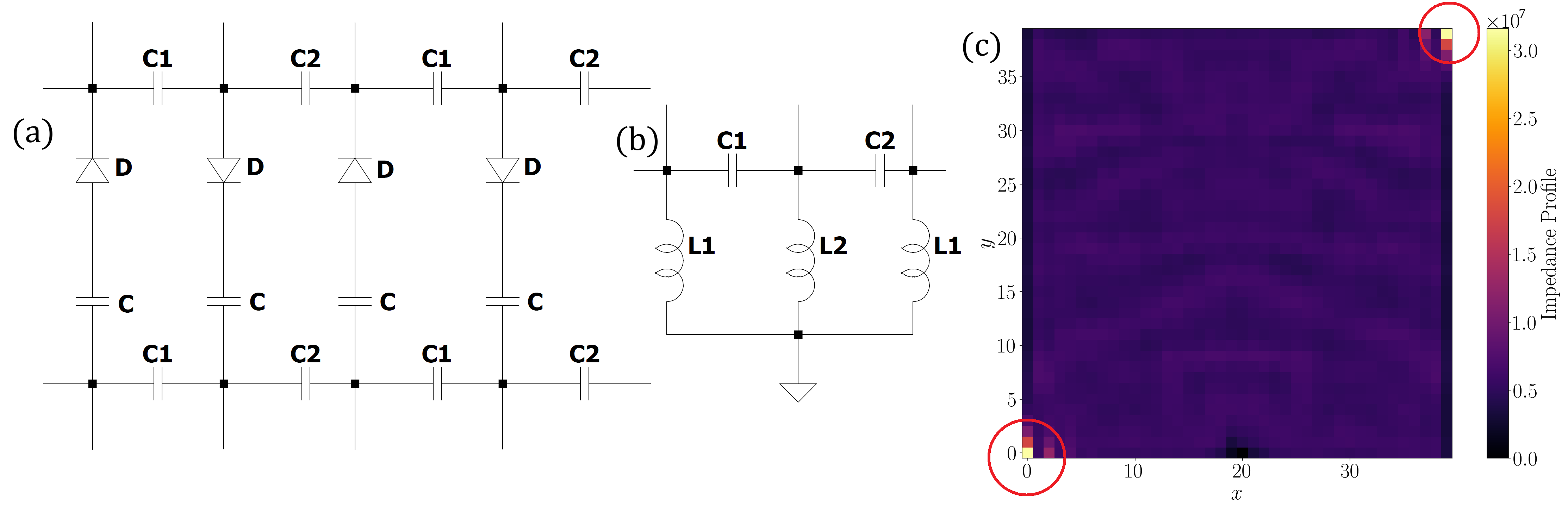}
\caption{(a) TEC design of NH square lattice. (b) Grounded inductors are associated with each node. (c) IP of the TEC, imitating the corner skin modes (yellow dots shown inside the red circles) in Fig.~\ref{Fig4}(d).}
\label{Fig5}
\end{figure*}
\noindent We now construct the TEC corresponding to the NH square lattice model.
The procedure involves substituting the atoms in a unit cell of the tight-binding (TB) model with nodes or junctions in the TEC.
Writing Kirchhoff's current and voltage laws for any random node, $i$, we shall have the following relation,
\begin{equation}
\mathcal{L}_{ij}= X_{ij} + \delta_{ij}W_i,
\quad\text{where}\quad 
W_i = \sum_{j} X_{ij} + X_i,
\label{eq:C3Lap}
\end{equation}
with $i = 1,\,2,\,3,\,\dots,2L_0.$
$X_{ij}$ is the conductance between two distinct nodes $i$ and $j$, and $L_0$ is number of total unit cells of the TEC.
$V_i$ and $I_i$ denote the voltage and the current flowing into node $i$ from the source placed elsewhere.
Note that the term $X_{ii}$ bears no meaning and can be set to zero, whereas $X_i$ is the resultant conductance between node $i$ and the ground.
This construction is relatively straightforward, since the system consists of SSH chains along the $x$-direction.
The only additional step is to couple these SSH circuits along the $y$-direction using unidirectional hoppings with amplitude $t$, as in the TB model.
To realize the unidirectional couplings along the $y$-direction, we employ diodes, which restrict the flow of the current along a single direction.
The complete TEC layout is shown in Fig.~\ref{Fig5}(a), while the configuration of the grounded inductors is depicted in Fig.~\ref{Fig5}(b).
The most convenient way to visualize the circuit's response is to measure the impedance between two nodes.
The impedance between two nodes of a circuit is given by the formula~\cite{Wu},
\begin{equation}\label{eq:Impedance}
Z_{ab} = \frac{V_a-V_b}{I_{ab}} = \sum_{j_n}\frac{|\phi_{n,a}-\phi_{n,b}|^2}{j_n}.
\end{equation}
where $|\phi_{n,a}-\phi_{n,b}|$ is the amplitude difference between $a$ and $b$ nodes of the $n^{\mathrm{th}}$ eigenmode, with $j_n$ being the $n^{\mathrm{th}}$ eigenvalue. 
The impedance measured between node $a$ and the ground is given by,
\begin{equation}
Z_{a,{\rm ground}}=\sum_{j_n}\frac{|\phi_{n,a}|^2}{j_n}.
\end{equation}
The above equation indicates that the impedance is highly sensitive to the wave-function distribution in the vicinity of $j_n=0$, as it diverges at this point in the absence of dissipation.
In realistic circuits, however, losses are unavoidable due to both the circuit components and the connecting wires.
Interestingly, these losses introduce a small imaginary contribution to the admittance spectrum, regularizing the divergence and rendering the impedance finite.
In experiments, the impedance can be measured by directly connecting the circuit nodes to the ports of a measurement instrument using wires.
In general, skin modes are visualized through the VP of the TEC~\cite{7t7k-qg49}.
However, for the SOSE, only $\mathcal{O}(L)$ skin modes exist out of a total of $\mathcal{O}(L^2)$ bulk modes.
In this situation, the HST modes are more effectively designated through the IP of the circuit.
This is because the corner skin modes exhibit a much stronger localization at the corners compared to bulk states, giving them a disproportionately large contribution to the impedance between two nodes, as described by Eq.~\eqref{eq:Impedance}, even though their energies are not exactly zero.
To measure the IP, the TEC is excited by injecting a current at a randomly chosen node, and the impedance at each node is recorded.
The angular frequency of the AC driving source is fixed at $\omega = 10^{6}\,\mathrm{MHz}$.
The circuit parameters are chosen as $C_1 = 0.5\,\mu\mathrm{F}$, $C_2 = 1\,\mu\mathrm{F}$, and $C = 0.5\,\mu\mathrm{F}$.
The grounded inductors are set to,
\begin{equation}
    L_1=L_2=\left[\omega^2\,(C+C_1+C_2)\right]^{-1}\approx0.67\,\mu\mathrm{H}.
\end{equation}
The resulting IP is shown in Fig.~\ref{Fig5}(c).
A pronounced enhancement of the impedance is observed at the two opposite corners, reaching values of order $\sim 3\times 10^{7}\,\Omega$.
This behavior faithfully reproduces the HST modes observed in the TB results shown in Fig.~\ref{Fig4}(d).

\subsection{\label{Sec3.3}SOSE in an NH BW Lattice}

\noindent We now turn to the NH BW lattice and examine the distinctive properties of the associated skin modes.
To introduce non-reciprocity into the otherwise Hermitian BW model, we render the hopping along the $y$-direction asymmetric in a certain patterned manner: two consecutive $y$-direction chains are made unidirectional with the same orientation. In comparison, the two subsequent chains are assigned unidirectional hopping in the opposite direction.
As we shall see, his staggered arrangement of non-reciprocity plays a crucial role in distinguishing the second-order skin modes obtained from the NH square lattice.
The bulk Hamiltonian for this model is given via,
\begin{align}
&H_{BW}(\Vec{k}) = \sum_{k_x,k_y} \Big[
\left(t_1+t_2e^{-ik_x}\right)c^{\dag}_{\Vec{k},A}c_{\Vec{k},B}\nonumber \\
&+ \left(t_1+t_2e^{ik_x}\right)c^{\dag}_{\Vec{k},C}c_{\Vec{k},D} + t\,c^{\dag}_{\Vec{k},A}c_{\Vec{k},D} + te^{ik_y}c^{\dag}_{\Vec{k},C}c_{\Vec{k},B}
\Big].
\label{eq:C7BSBW}
\end{align}
Since the BW lattice is obtained by selectively removing hoppings from the square lattice, it does not possess mirror symmetry about the $x$-axis, that is, it lacks $\mathcal{M}_x$.
It does, however, retain mirror symmetry about the $y$ axis, with the corresponding operator given by $\mathcal{M}_y \equiv \tau_x \sigma_x$.
Since inversion can be viewed as the product of the two mirror operations, $\mathcal{M}_x$ and $\mathcal{M}_y$, the Hermitian BW lattice does not exhibit inversion symmetry.
As in the previous case, the Hamiltonian $H_{BW}(k)$ cannot be promoted to a Hermitian Hamiltonian that preserves inversion symmetry and supports intrinsic second-order topology.
Therefore, the second-order skin modes realized in the NH BW lattice are also extrinsic in nature.
In other words, they correspond to HST modes, arising from the interplay between non-hermiticity and lower-dimensional topological features rather than from an intrinsic bulk second-order topology.

\begin{figure}
\centering
\includegraphics[width=0.5\textwidth]{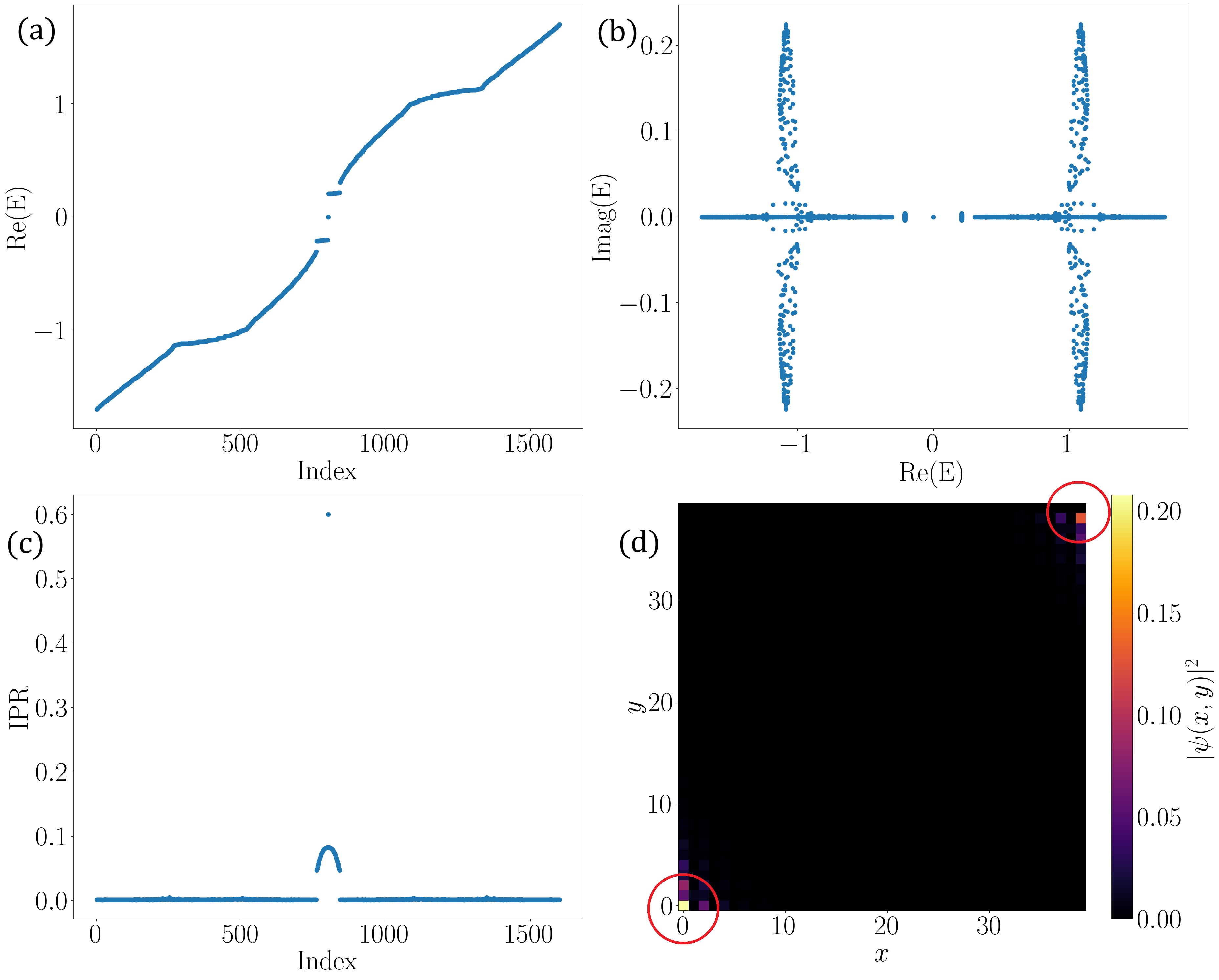}
\caption{(a) The real part of the energy spectra of an NH BW lattice with OBC in both directions for $t<t_1+t_2$. (b) The spectral distribution in the complex plane. The corner skin modes now form EP-like structures. (c) IPR of the corner skin modes is non-zero. However, two zero-energy modes have IPRs much higher than those of the corner skin modes. (d) The probability distribution of a random second-order skin mode is shown via yellow dots inside the red circles.}
\label{Fig6}
\end{figure}

Now we turn to the most interesting part of this manuscript.
The NH BW lattice exhibits substantially richer physics than the NH square lattice.
As shown in Fig.~\ref{Fig6}(a), the real part of the spectrum develops a clear energy gap.
The small step-like structures visible inside this gap correspond to the HST modes.
The concept of the HST effect was first proposed in 2019 by Lee et. al. ~\cite{PhysRevLett.123.016805}, based on a 2D extension of the 1D NH SSH model exhibiting NHSE.
In this construction, the model is formed by stacking the copies of the 1D system along a spatial direction and is characterized by four independent non-reciprocal parameters.
When the non-reciprocities are balanced, destructive interference occurs along one direction, giving rise to a line NHSE.
By contrast, if the non-reciprocities cancel along both directions, the bulk states no longer exhibit NHSE along either axis; nevertheless, a corner NHSE remains under OBC.
This residual localization arises from first-order topological states that are unevenly distributed across sublattices.
Such states do not undergo complete destructive interference due to non-reciprocity, allowing them to retain NHSE and thereby form a new class of higher-order skin states.

In addition, the spectrum in Fig.~\ref{Fig6}(a) contains two zero-energy edge modes that originate from the parent Hermitian BW lattice.
To be precise, in a Hermitian BW lattice of size $L_x \times L_y$, there are $2L_y$ first-order topological edge modes for $t \le t_1$.
These originate from the $L_y$ independent SSH chains along the $x$-direction, with each chain contributing two topological edge modes.
Among these $2L_y$ modes, however, only two lie exactly at zero energy ($E=0$).
The remaining $(2L_y - 2)$ modes have finite energies $E = \pm t$.
Importantly, the two zero-energy modes are localized at two corners of the lattice, whereas the other $(2L_y - 2)$ first-order modes are localized along the $y$ edges.
These $(2L_y - 2)$ modes are degenerate at energies $E=\pm t$ in the Hermitian limit, but this degeneracy is not associated with the EPs.
\begin{figure*}
\centering
\includegraphics[width=\textwidth]{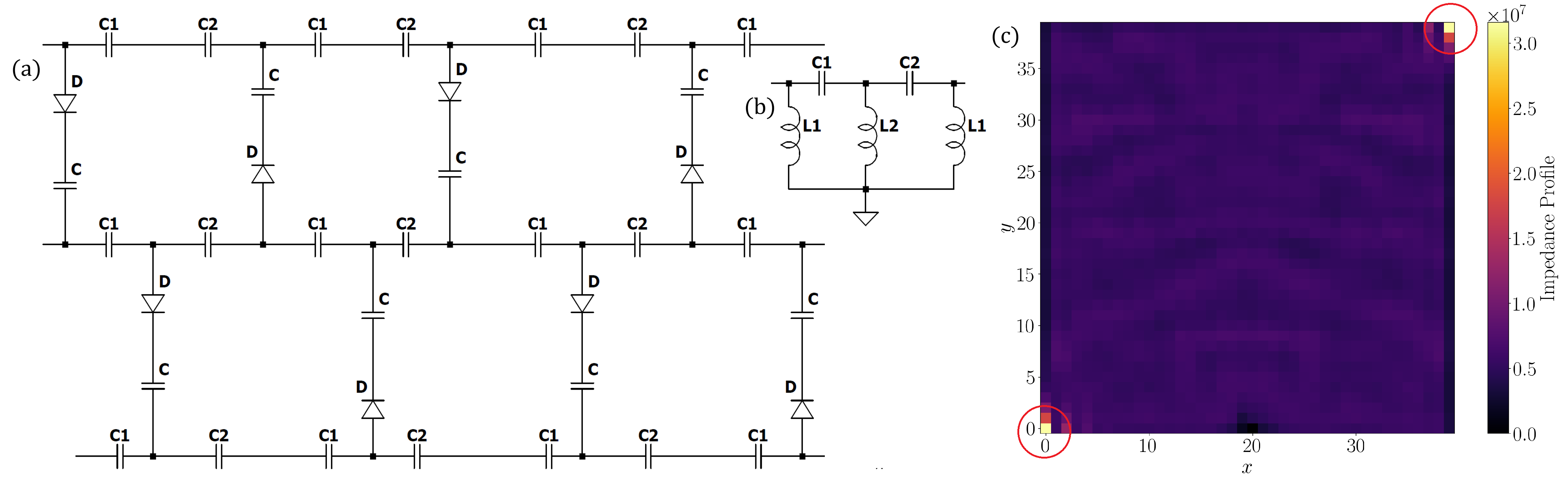}
\caption{(a) TEC design of NH BW model. (b) Grounded inductors, associated with each node, are shown. (c) IP of the TEC, imitating the corner skin modes, shown via yellow dots inside the red circles, in Fig.~\ref{Fig6}(d).}
\label{Fig7}
\end{figure*}
Upon introducing non-hermiticity, these $(2L_y - 2)$ degenerate edge modes evolve into extrinsic second-order skin modes localized at the other two corners of the lattice.
Their energies become almost degenerate, acquiring only a very small imaginary part, which gives rise to the step-like features observed in Fig.~\ref{Fig6}(a).
These modes are neither zero-energy modes nor part of the bulk spectrum; instead, they remain spectrally isolated from the bulk bands.
In Fig.~\ref{Fig6}(b), these modes appear as two point-like clusters in the complex-energy plane.
These skin modes are remarkable for two key reasons: (i) They originate from degeneracies of the parent Hermitian Hamiltonian (the Hermitian BW lattice in this case). Consequently, despite their appearance, they are not exceptional points. (ii) They are almost dynamically stable, possessing only negligible imaginary parts in their energies. This behavior is highly unusual compared to previously studied HST systems~\cite{PhysRevLett.123.016805, PhysRevB.102.241202, PhysRevB.102.205118}, where the skin modes typically exhibit appreciable amplification or decay.
Fig.~\ref{Fig6}(c) shows the IPR of the corner skin modes.
$(2L_y-2)$ modes have a lower IPR, while the remaining two zero-energy modes inherited from the Hermitian BW lattice become localized at a single corner of the lattice.
Fig.~\ref{Fig6}(d) shows the probability distribution of a random second-order skin mode, localized at two opposite corners of the lattice.

\subsection{\label{Sec3.4}TEC of an NH BW Lattice}

\noindent We conclude this discussion by constructing the TEC corresponding to the NH BW model.
The procedure closely follows that used for the NH square lattice in Section~\ref{Sec3.2}.
The only modifications required are the removal of selected capacitors along the $y$-direction and an appropriate reorientation of the diodes along the $y$-axis, which together implement the NH BW lattice geometry and its non-reciprocal couplings.
The resulting NH BW circuit diagrams are shown in Figs.~\ref{Fig7}(a) and \ref{Fig7}(b).
All the circuit element values are kept identical to those used in Section~\ref{Sec3.2}.
The corresponding IP is presented in Fig.~\ref{Fig7}(c).
This IP closely resembles that of the NH square lattice shown in Fig.~\ref{Fig5}(c) and successfully reproduces the HST behavior observed in the TB results of Fig.~\ref{Fig6}(d).
It should be noted, however, that the two special corner skin modes identified in the NH BW TB model cannot be individually resolved in the TEC realization.
This limitation arises because, in terms of the probability density, these modes are indistinguishable from the other corner-skin modes when computing the impedance via Eq.~\eqref{eq:Impedance}.
As a result, while the TEC faithfully captures the overall HST localization, it does not allow separate identification of these particular corner modes.

\section{\label{Sec4}Conclusion}

\noindent We have investigated the NH variant of the BW model, which realizes an HST effect (or extrinsic SOSE), and systematically compared its behavior with that of the NH square lattice.
To set the stage, we first introduced the Hermitian BW model and analyzed its spectral properties.
In the Hermitian variant, the BW lattice does not host intrinsic higher-order topology; any topological features arise solely from the dimerized hopping, which emulates SSH chains along the $x$-direction.
Upon introducing non-hermiticity, the BW model displays several unusual features in the spectral properties of the HST modes.
In contrast to previously studied SOSE systems, these modes do not generate non-trivial spectral windings in the complex plane.
Instead, they appear similar to the EPs in the complex-energy spectrum, although they are not genuine EPs.
Among all the corner skin modes, only two, which descend from the topological corner modes of the Hermitian BW lattice, are localized at individual corners.
The remaining corner skin modes accumulate at the opposite pair of corners.
This behavior is markedly different from that observed in the NH square lattice, where the corner skin modes are distributed more uniformly.
For both the NH square lattice and the NH BW lattice, we have constructed corresponding TEC realizations.
These circuit implementations enable direct visualization of the HST modes and faithfully simulate an experimental platform, thereby corroborating the theoretical predictions.

\bibliography{ref7.bib}
\end{document}